\documentclass[11pt]{article}

\usepackage{a4}
\usepackage{amssymb}
\usepackage{amsmath}

\usepackage[latin1]{inputenc}
 
\newcommand{\lneg}                  {{\mathop{\neg}}}
\newcommand{\limp}                  {\mathbin{\supset}}
\newcommand{\leqv}                  {\mathbin{\equiv}}
\newcommand{\lconj}                 {\mathbin{\wedge}}
\newcommand{\ldisj}                  {\mathbin{\vee}}
\newcommand{\lneqv}               {\mathbin{\not\equiv}}

\begin{document}
 
\title{Epistemic nature of quantum reasoning}

{\author{Alfredo B. Henriques $^{\small(1)}$ and Amílcar Sernadas $^{\small(2)}$\\[1mm]
{\scriptsize $^{\small(1)}$ Centro Multidisciplinar de Astrofísica and Departamento de Física}\\[-1mm]
{\scriptsize $^{\small(2)}$ SQIG, Instituto de Telecomunicações and Departamento de Matemática}\\
{\scriptsize Instituto Superior Técnico, Universidade de Lisboa, Portugal}\\[-1mm]
{\scriptsize \{alfredo.henriques,amilcar.sernadas\}@tecnico.ulisboa.pt}}

\date{April 21, 2015}

\maketitle

\begin{abstract}
Doubts are raised concerning the usual interpretation of the alleged failure, by quantum mechanics, of the distributive law of classical logic. The difficulty raised by incompatible sets of observables is overcome within an epistemic enrichment of classical logic that provides the means for distinguishing between the value of a variable and its observation while retaining the classical connectives.
\\[2mm]
\end{abstract}


Since the seminal work of von Neumann and Birkhoff, in the 1930's (see~\cite{bir:neu:36}), it is commonly assumed (see Putnam~\cite{put:68}) that the experimental propositions of quantum mechanics can break the distributive law of classical logic. Due to the importance of this statement to the construction and rise of quantum logic, we re-examine a simple, but rather typical, example borrowed from~\cite{wiki}. 

Consider a particle moving along a line, taken as $x$-axis, and write the following propositions (we use the so called natural units, taking the velocity of light $c=1$ and $\hslash=1$):
\begin{itemize}
\item p = ``the particle has linear momentum $p_x$ in the interval $[0, 1/6]$";
\item q =``the particle is in the interval on the $x$-axis $[-1, 1]$";
\item r =``the particle is in the interval on the $x$-axis $[1, 3]$".
\end{itemize}

In this system of units, the uncertainty principle is expressed as
$$\Delta p_x\Delta x \geq 1/2.$$
The distributive law of propositional logic is given by the expression
$$p \lconj (q \ldisj r) \leqv (p \lconj q) \ldisj (p \lconj r).$$
Consider the usual reasons given for the failure of this law. The left hand side (lhs) is read as telling us that the particle, with momentum in the interval $[0,1/6]$, has position in the interval $[-1, 3]$, compatible with the uncertainty principle: $\Delta p_x= 1/6$ and $\Delta x=4$, their product being 
$2/3 > 1/2$. Thus, it has been argued that $p \lconj (q \ldisj r)$ is true. Rigorously, one should say that $p \lconj (q \ldisj r)$ can be 
true (depending on the truth values of p, q and r).
As for the right-hand-side (rhs), we immediately see that $p \lconj q$ cannot be true, as the product of the uncertainties is $\Delta p_x\Delta x = 1/3 < 1/2$. The same applies to the second term, where, again, we have $\Delta p_x\Delta x = 1/3 < 1/2$. The usual conclusion is that the rhs 
$(p \lconj q) \ldisj (p \lconj r)$ is false. Rigorously, one should say that $(p \lconj q) \ldisj (p \lconj r)$ must always be false. Hence, the distributive law, valid in classical logic, can fail when dealing with quantum mechanics.

We now look more carefully at what has been argued.

With the intervals defined as above, we immediately see that, when considering propositions $p \lconj q$ and $p \lconj r$, we are, in fact, considering propositions which violate the uncertainty principle, that is, they break basic rules of quantum mechanics, with the consequence that they cannot express experimental/testable propositions. We cannot perform experiments allowing us to make simultaneous measurements of the values of the linear momentum and the position within those intervals; they correspond to mutually exclusive experimental situations. If we took them as being made simultaneously, we would be making the sin of a counterfactual reasoning: we would be assuming as being observed something which in fact has not been observed since it cannot be observed at all.

To show that quantum theory breaks some established rule of classical logic, like the distributive property, are we not supposed to argue in a way consistent with the theory itself and should we not avoid using counterfactual reasoning?

Is it appropriate, in any reasonable interpretation of the expressions of classical logic, to take cases and situations - perfectly possible in classical physics, but not within quantum mechanics - and use them to claim that quantum mechanics breaks some rule of classical logic and, then, using this to claim that logic is empirical?

Measurements actually performed would certainly give us values in agreement with the uncertainty principle. 
For example, if the intervals in the propositions q and r were slightly larger, as would result from actual measurements, then, either $p \lconj q$ or $p \lconj r$ could be true and the distributive law would be valid.

The same problem can be formulated when we use the more sophisticated language of the Hilbert spaces, for instances in Bacciagaluppi (see~\cite{bac:09} also \cite{put:68}). The equivalent to the counterfactual error above is made, although in a more rigorous setting.

Other examples could be found, like those based on the Young interference observations, where, again, the same kind of incorrect use is made of incompatible observations.

To advance a little more with our analysis, we go back to our example and construct the truth table corresponding to the propositions p, q, r. We have the possibilities depicted in Figure~\ref{fig:tt}. Sign (*) marks those table entries that correspond to incompatible values of the propositions, corresponding to the mutually exclusive experimental situations mentioned above; these incompatibilities mean that we cannot even complete the truth table for the distributive law. 

Can we, then, use the connectives of classical logic, without the truth table being complete? Are we then allowed to talk about the distributive law? Definitely not, at least according to~\cite{dum:76}.

\begin{figure}[ht] 
$$
\begin{array}{|c|c|c|c|c|c|}
\hline
 & p & q & r & p\lconj (q\ldisj r) & (p\lconj q)\ldisj (p\lconj r)\\
\hline
 & 0 & 0 & 0 & 0 & 0\\
\hline
 & 0 & 0 & 1 & 0 & 0\\
\hline
 & 0 & 1 & 0 & 0 & 0\\
\hline
 & 0 & 1 & 1 & 0 & 0\\
\hline
 & 1 & 0 & 0 & 0 & 0\\
\hline
\ast & 1 & 0 & 1 & \times & \times\\
\hline
\ast & 1 & 1 & 0 & \times & \times\\
\hline
\ast & 1 & 1 & 1 & \times & \times\\
\hline
\end{array}
$$
\caption{Truth table for $p\lconj (q\ldisj r)$ and $(p\lconj q)\ldisj (p\lconj r)$.}\label{fig:tt}
\end{figure}

There are, however, a few additional critical points that we want to address. 
When we write $p$, $q$ or $r$, we take it as meaning that those measurements have been made, and that we know the corresponding results. Once again, remember these measurements are assumed to be made simultaneously, meaning that $p \lconj (q \ldisj r)$ is not a legitimate expression, as we have in fact two determinations, two measurements, of the position, each one of them incompatible with the measurement of the momentum, given the intervals previously defined. If it is to be a valid expression, we must refer to a single measurement of position in the interval $[-1,3]$. This suggests that there has been an initial bad choice of propositions. What we need is a proposition like 
$$s = \text{``the particle was observed in the interval on the $x$-axis} \, [-1, 3]"$$
which, according to quantum mechanics, is not the same thing as saying ``the particle was observed in the interval $[-1, 1]$ or was observed in the interval $[1, 3]$".

Nevertheless, one may want to work with possibly incompatible propositions. In such a scenario, two possibilities seem to be open to us. Either we accept that the connectives may have different properties from those in classical logic, and we work with a different logic (adopting the stance first proposed in ~\cite{bir:neu:36}), or we may instead try to retain the classical connectives, as they are, and adopt an epistemic approach as we proceed to explain.

In order to overcome the difficulties mentioned above, raised by incompatible observables, we propose to extend classical logic with the means for distinguishing between, for instance, ``linear momentum is in [0, 1/6]" and ``linear moment was measured to be in [0, 1/6]". To this end, if we introduce the epistemic operator $K$ (see~\cite{hin:62,fag:95,mey:hoe:95}), for knowing (after observing), we can express that the knowledge of $p$ is incompatible with the knowledge of $q$ and with the knowledge of r:
$$K(p) \limp \lneg K(q)$$
and 
$$K(p) \limp \lneg K(r).$$
Hence, $K(p) \lconj (K(q) \ldisj K(r))$ must always be false, while $K(p \lconj s) \leqv K(p) \lconj K(s)$ may be true. 

Observe that the law $K(a \lconj b) \leqv K(a) \lconj K(b)$ is characteristic of an epistemic operator (see, for instance, Chapter 1 of \cite{mey:hoe:95}). 
On the other hand, one should not expect $K(a \ldisj b) \leqv K(a) \ldisj K(b)$ to hold in general (ibidem). 
Hence, the epistemic $K$ does reflect the quantum physics reality: 
$$K(q \ldisj r) \lneqv K(q) \ldisj K(r),$$
that is, observing that the position is in $[-1, 3]$ is not equivalent to observing that it is in $[-1, 1]$ or to observing that it is in $[1, 3]$.

Therefore, according to the properties of the epistemic operator $K$ (known = quantum observed), there is no surprise at all in 
$$K(p \lconj (q \ldisj r)) \lneqv K(p \lconj q) \ldisj K(p \lconj r).$$
Indeed, while 
$$K(p \lconj (q \ldisj r)) \leqv K(p) \lconj K(q \ldisj r)$$
and 
$$K(p \lconj q) \ldisj K(p \lconj r) \leqv (K(p) \lconj K(q)) \ldisj (K(p) \lconj K(r)) \leqv K(p) \lconj (K(q) \ldisj K(r))$$
because $K$ distributes over conjunction, it does not distribute over disjunction and, so, $K(q \ldisj r)$ is not necessarily equivalent to $K(q) \ldisj K(r)$.

Recall that none of the difficulties we have discussed appear when we describe the state of the quantum system using what is called a complete set of compatible observables, whose operators commute among themselves, giving a complete description of the state of the system (see, for instance, Section 13 of Chapter II in \cite{dir:58} or Section 1 of Chapter 1 in \cite{lan:lif:66}). These observables can, in principle, be precisely and simultaneously measured, and no problems arise with the distributive law. This is to be expected, as quantum mechanics is based on a ``mathematical formalism that is, itself, based firmly on a classical two-valued logic" (see~\cite{ish:01}). In such a scenario, $K(a)=a$, since we do not need to distinguish between them, and the epistemic enrichment of classical logic would collapse back into classical logic.

After all that has been said, can we still claim that logic is empirical, by appealing to quantum mechanics? Quantum mechanics is very different from classical physics, but does not break classical logic. 
It just shows that classical logic is not rich enough. In our opinion, given the epistemological constraints imposed by quantum mechanics on what can be measured, the missing ingredient is necessarily of an epistemic nature.


\section*{Acknowledgments}

This work was was partially supported by the PQDR initiative of SQIG at Instituto de Telecomunicações, 
Fundação para a Ciência e a Tecnologia via projects UID/FIS/00099/2013 and UID/EEA/50008/2013,
and the European Union through project GA 318287 LANDAUER.


\end{document}